\documentclass[prd, nobalancelastpage, secnumarabic,amsmath,amssymb,nofootinbib,hyperref=pdftex,11pt]{revtex4}

\usepackage{mathtools}
\usepackage{amssymb}
\usepackage{amsmath}
\usepackage{float}
\usepackage{gensymb}
\usepackage{siunitx}
\usepackage{caption}
\usepackage{xcolor}
\usepackage{soul}
\usepackage{subcaption}

\usepackage{chapterbib}    
\usepackage{color}         
\usepackage{graphics}      
\usepackage[pdftex]{graphicx}      
\usepackage{epsf}          
\usepackage{bm}            
\usepackage{verbatim}			
\usepackage[colorlinks=true]{hyperref}
\begin{document}

\title{Quantum Instrumentation Control Kit – Defect Arbitrary Waveform Generator (QICK-DAWG): A Quantum Sensing Control Framework for Quantum Defects}
\author{E. G. Riendeau$^{1,2}$, L. Basso$^2$, J. J. Mah$^2$, R. Cong$^{2,3}$, M. A. Sadi$^{2,4}$, J. Henshaw$^2$, K. M. Azizur-Rahman$^2$, A. Jones$^{5,2}$, G. Joshi$^2$,  M. P. Lilly$^2$, and A. M. Mounce$^2$ }

\affiliation{$^1$Haverford College, $^2$Center for Integrated Nanotechnologies, Sandia National Laboratories, $^3$Brown University, $^4$Purdue University, $^5$Georgia Institute of Technology}

\date{29 November 2023}

\begin{abstract}
    Quantum information communication, sensing, and computation often require complex and expensive instrumentation resulting in a large entry barrier.  The Quantum Instrumentation Control Kit (QICK) overcomes this barrier for superconducting qubits with a collection of software and firmware for state-of-the-art radio frequency system on chip (RFSoC's) field programmable gate architecture (FPGA) chips.  Here we present a software and firmware extension to QICK, the Quantum Instrumentation Control Kit - Defect Arbitrary Waveform Generator (\href{https://github.com/sandialabs/qick-dawg}{QICK-DAWG}), which is an open-source software and firmware package that supports full quantum control and measurement of nitrogen-vacancy defects in diamond and other quantum defects using RFSoC FPGAs. QICKDAWG extends QICK to the characterization of nitrogen-vacancy defects and other diamond quantum defects by implementing DC-1 GHz readout, AOM or gated laser control, and analog or photon counting readout options. QICK-DAWG also adds pulse sequence programs and data analysis scripts to collect and characterize photoluminescence (PL) intensity, optically detected magnetic resonance (ODMR) spectra, PL readout windows, Rabi oscillations, Ramsay interference spectra, Hahn echo spin-spin relaxation times T$_2$, and spin-lattice relaxation times T$_1$. We demonstrate that QICK-DAWG is a powerful new paradigm of open source quantum hardware that significantly lowers the entry barrier and cost for quantum sensing using quantum defects.
\end{abstract}
\maketitle

\section{Summary}	
The Quantum Instrumentation Control Kit - Defect Arbitrary Waveform Generator (QICK-DAWG), is an open-source software and firmware package for full quantum control and measurement of defect centers in diamond and other semiconductor materials for quantum sensing \cite{QICK-DAWG}. QICK-DAWG extends the capabilities of the Quantum Instrumentation Control Kit (QICK, an open-source qubit firmware and software package) to defect centers by implementing controlled laser pulsing and low frequency readout required for defect initialization, control, and measurement using recently available Radio Frequency System-on-Chip (RFSoC) Field Programmable Gate Arrays (FPGA). In addition to user-friendly software and firmware, QICK-DAWG adds documentation that walks users through modifying the hardware of commercial off-the-shelf evaluation boards (from Real Digital and Xilinx) to digitize low frequency signals. QICK-DAWG’s measurement programs consist of specific microwave, laser, and readout pulse sequences built in a Python framework for consistency and extensibility. Pulse sequence programs and data analysis scripts are included to collect and characterize photoluminescence (PL) intensity, optically detected magnetic resonance (ODMR) spectra, PL readout windows, Rabi oscillations, Ramsay interference spectra, Hahn echo spin-spin relaxation times T$_2$, and spin-lattice relaxation times T$_1$. Additional pulse sequence programs and data analysis scripts will be added in the future. QICK-DAWG also implements live-update of plots for PL intensities to optimize laser alignment, broadband ODMR spectra for magnetic field alignment, and Rabi oscillations to optimize microwave antenna alignment or positioning. 

The QICK-DAWG package consists of FPGA firmware (modified from QICK firmware) compiled in Vivado, Python software that extends QICK for specific pulse programs and altered functionality, instructions for installation and hardware modifications, and a demo Jupyter notebook. The Readme.md file walks users through rudimentary hardware setup, installation, and modification required for low frequency data collection using RealDigital’s RFSoC4x2. The package also has a batch file with directions for easy installation setup for the RFSoC Linux kernel that drives firmware, controls, and reads from the FPGA.  The demo Jupyter Notebook walks users through typical experimental flow and the configuration of each pulse sequence. Each measurement program includes a method that checks whether all required configurations for pulsing programs are present and a method which provides a visual representation of the plot sequence. Ultimately, QICK-DAWG is an extensible firmware and software package for quantum sensing using NVs and other quantum defects using one control hub for consistency among different experimental setups or laboratories.  

\section{Related Work}

Open-source software packages for quantum control and data acquisition such as ARTIQ\cite{Bourdeauducq:2016}, QubiC \cite{Xu:2023}, and QICK \cite{Stefanazzi:2022} have been continually developed over the past decade for a variety of quantum experiments. ARTIQ--Advanced Real-Time Infrastructure for Quantum Physics--began in 2013 at the NIST ION Storage Group \cite{Bourdeauducq:2016}. ARTIQ was created in response to the shortcomings in both in-house and existing commercial based quantum control systems. ARTIQ helps describe complex experiments and runs on a high-level Python based programming language and is executed on an FPGA. ARTIQ has since partnered with the hardware project Sinara to improve quality and reproducibility of the hardware being used \cite{Kulik:2022}. The ARTIQ/Sinara software-hardware combination is focused on the fast control of ion-based quantum processors and has been used in nearly one thousand physics experiments \cite{Shammah:2023}.  

Focusing on hardware, Xilinx’s ZCU216, Xilinx’s ZCU111 and Real Digital’s RFSoC4x2 are recently available RFSoC FPGAs with the ability to generate control pulses at high frequencies (6–10GHz) and to digitize signals from photodiodes and single photon detector modules at high sample rates. The precise high frequencies pulse generation, readout capability, compact size, and relatively low cost of these RFSoC FPGAs make them ideal candidates for defect control hardware. There are two open-source quantum hardware software packages currently supporting the use of RFSoC’s: Qubic and QICK. 

QubiC, started in 2018 at the Advanced Quantum Testbed at the Lawrence Berkeley National Laboratory, is an FPGA based open-source control and acquisition software, and it is focused on superconducting qubits. Recently, QubiC adopted Xilinx's ZCU216 to take advantage of new RFSoC capabilities \cite{Xu:2021}\cite{Xu:2023}. 

QICK (Quantum Instrumentation Control Kit) is the second package to utilize RFSoC FPGA’s, developed in 2021 \cite{ Stefanazzi:2022}. QICK has been applied to superconducting, spin, atomic, molecular, and optical qubit systems, and has reached 40 labs in the last two years \cite{Shammah:2023}. Most appealing for quantum sensing, QICK has complied firmware for Real Digital’s RFSoC4x2 which is the lowest cost option of commercial off the shelf FPGA boards.

Overall, QICK provides a sensible foundation to build upon to produce an open-source control measurement, software, and firmware package for NVs and other diamond quantum defects as QICK provides specific firmware for the RFSoC4x2, ZCU111, and ZCU216—firmware not available in the other quantum control FPGA packages. QICK provides firmware, a high-level Python based user interface, and accessible inexpensive hardware—RFSoC FPGAs. However, simple modification to QICK firmware, hardware, and software was necessary for implementation of QICK in spin-based quantum control experiments like NVs and other diamond quantum defects. Firmware modification is required to sync clock registers, rudimentary hardware modification is necessary to support low frequency readout, and software modification adds specific coded pulse sequences to support commonly used quantum control and characterization methods. The RFSoC4x2 requires the user to bypass a high pass balun to access low frequency digitization. Nonetheless, QICK-DAWG and the RFSoC4x2 create an extremely low barrier of entry for open-quantum hardware (OQH) with accessible API, and with the modifications made by QICK-DAWG is suitable for spin-based experiments such as NVs and other defects in diamond.

\section{Statement of need}
OQH is a broad category that covers the open-source tools and components needed to build and control quantum computers, technologies, and sensors. OQH has the potential to accelerate quantum research, quantum technology development, and increase accessibility of quantum computing and quantum sensing. The specific area of OQH that QICK-DAWG fills is open-source instrumentation, control, and data acquisition software.

Currently, quantum sensing with defects is accomplished by research groups assembling their own hardware consisting of many different instruments and creating their own software to control them \cite{Butcher:2019}. Timing between instruments, such as microwave generators and switches, arbitrary waveform generators, digital to analog converters, and/or photon counters, is typically provided by a fast FPGA TTL generator triggering independent instruments which can lead to timing offsets that must be accounted for.  Thus, there is a lack of consistency in hardware and software for quantum defect-based quantum sensing, a niche that QICK-DAWG fills, in addition to simplified operation as all functionality happens on a single instrument’s timing.

QICK-DAWG   increases accessibility to NV and defect research through its open-source nature, reduced hardware cost, high-level Python based user interface, and extensive documentation. In the included documentation and demo, the QICK-DAWG package utilizes the Real Digital' RFSoC4x2 as this particular SoC has a relatively low cost. Thus, QICK-DAWG lowers the cost of entry for quantum sensing with RFSoCs by as much as an order of magnitude. QICK-DAWG implementation requires some rudimentary hardware modification and additional hardware that is detailed in the installation Readme document. Thus, QICK-DAWG supports full quantum control and data acquisition of NV centers and other diamond quantum defects—a spin-based open-source software niche not yet filled. Furthermore, through reduced equipment costs, high-level Python user interface, and extensive documentation, QICK-DAWG makes NV quantum measurements accessible to academic and small labs and does not require extensive background knowledge in quantum control hardware. 

\section{Example Features}

QICK-DAWG has standardized measurement programs for acquiring PL intensity, ODMR, PL readout windows, Rabi oscillations, Hahn echo T$_2$, and spin-lattice relaxation time T$_1$. Each of these measurement programs:
1.	Checks that all required experimental configuration settings are present.
2.	Generates the assembly language code to generate the correct pulse sequence. 
3.	Executes the pulse sequence microwave signal, laser control, and data collection.
4.	Analyzes data returned from the FPGA.
5.	Graphically displays the pulse sequence, labeled with either configuration variable names or values.

\begin{figure} [H] 
    \centering
    \includegraphics[width=4.5in]{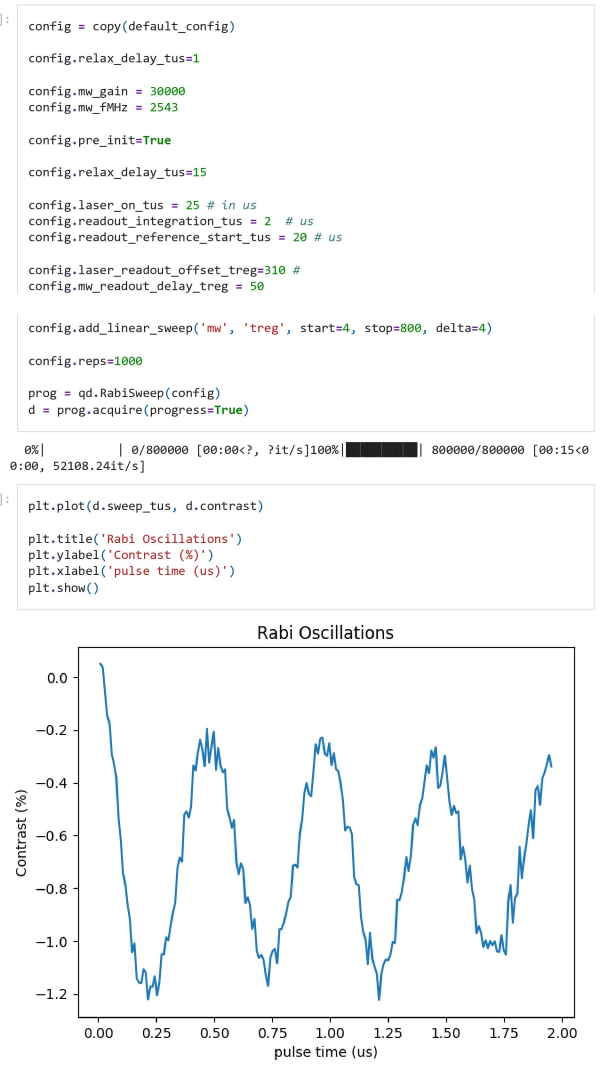}
    \caption{Example use of QICK-DAWG to perform Rabi oscillations from the demo Jupyter notebook.}
    \label{fig:Rabi}
\end{figure}

An example configuration, execution, and data plotting of our Rabi Oscillation program is displayed in Figure \ref{fig:Rabi}, with configuration-value labeled pulse sequence graphic. 

\begin{figure} [H] 
    \centering
    \includegraphics[width=4.25in]{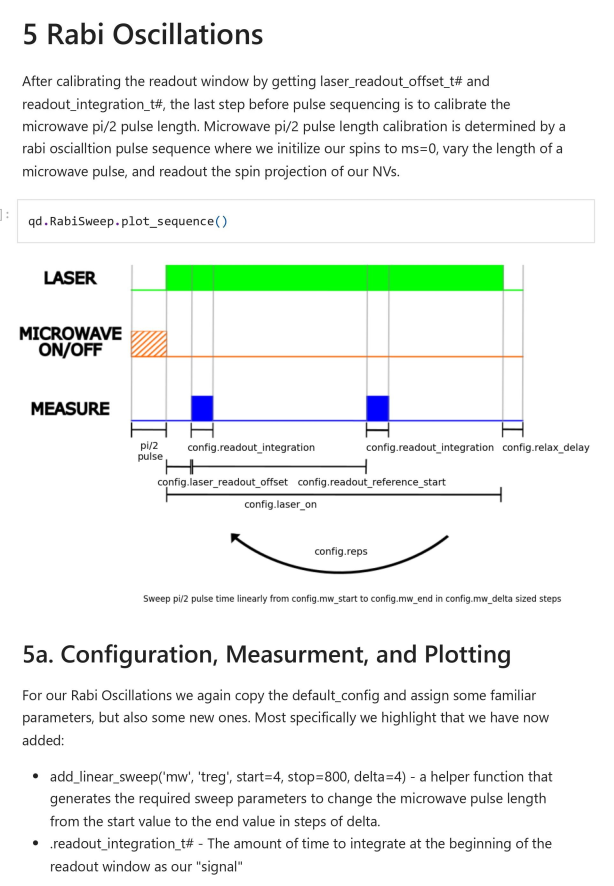}
\end{figure}

\section{Current Implementation}

At Sandia National Laboratories, we are currently using QICK-DAWG to characterize ensembles of both NVs in diamond and boron-vacancy defects in hexagonal boron nitride. These quantum defects are characterized for PL intensity, T$^{*}_2$, T$_2$, and T$_1$.  We characterize both the intrinsic properties of these quantum defects and dependent properties which change in the presence of other materials of interest, i.e., in a quantum sensing experiment. These methods allow us to understand the spin properties of low dimensional materials which cannot be accessed by traditional spin probes \cite{Henshaw:2023,Henshaw:2022,Casola:2018,Wang:2022}.

\section{Acknowledgements}

We thank S. Uemura and S. Sussman for helping to modify QICK firmware and guidance on using QICK. We would also like to acknowledge J. Feder, J. Heremans, C. Yao, E. Henriksen, F. Balakirev, and J. Kitzman for inspirational and technical discussions. 

Sandia National Laboratories is a multi-mission laboratory managed and operated by National Technology and Engineering Solutions of Sandia (NTESS), LLC, a wholly owned subsidiary of Honeywell International, Inc., for the DOE’s National Nuclear Security Administration under contract DE-NA0003525. This work was funded, in part, by the Laboratory Directed Research and Development Program and performed, in part, at the Center for Integrated Nanotechnologies, an Office of Science User Facility operated for the U.S. Department of Energy (DOE) Office of Science. This paper describes objective technical results and analysis. Any subjective views or opinions that might be expressed in the paper do not necessarily represent the views of the U.S. Department of Energy or the United States Government.

This work was supported in part by the U.S. Department of Energy, Office of Science, Office of Workforce Development for Teachers and Scientists (WDTS) under the Science Undergraduate Laboratory Internships Program (SULI). This work was performed, in part, at the Center for Integrated Nanotechnologies (CINT), an Office of Science User Facility operated for the U.S. Department of Energy (DOE) Office of Science.

\bibliography{main}
\end{document}